\documentclass[letterpaper,11pt]{article}
\pdfoutput=1 
             
\usepackage{jheppub} 

\usepackage[T1]{fontenc}
\usepackage{graphicx}
\usepackage{amsmath}
\usepackage{amssymb}
\usepackage{hyperref}
\usepackage[table]{xcolor}
\usepackage{bm}

\newcommand{\GeV}{\textrm{GeV}}

\preprint{LCTP-23-10}
\title{Precision Electroweak Tensions and a Dark Photon}

\author[a,b,c]{Keisuke Harigaya,}
\author[d]{Evan Petrosky,}
\author[d]{Aaron Pierce}

\affiliation[a]{Department of Physics, University of Chicago, Chicago, IL 60637, USA}
\affiliation[b]{Enrico Fermi Institute and Kavli Institute for Cosmological Physics, University of Chicago, Chicago, IL 60637, USA}
\affiliation[c]{Kavli Institute for the Physics and Mathematics of the Universe (WPI),
The University of Tokyo Institutes for Advanced Study,
The University of Tokyo, Kashiwa, Chiba 277-8583, Japan}
\affiliation[d]{Leinweber Center for Theoretical Physics, Department of Physics, University of Michigan, Ann Arbor, MI 48109, USA}

\emailAdd{kharigaya@uchicago.edu}
\emailAdd{epetros@umich.edu}
\emailAdd{atpierce@umich.edu}

\abstract{We examine how different assumptions about the hadronic vacuum polarization, the $W$ boson mass, and the forward-backward asymmetry in $b$-quarks at the $Z$ pole can impact the precision electroweak fit.  We study the implications for a kinetically mixed dark photon,  addressing the complementarity of precision bounds and direct searches, particularly in the case where the dark photon can decay into the dark sector, and we consider implications for future Large Hadron Collider searches.   We comment on cases where the precision effects of the dark photon may not be well-described by the oblique parameters. }

\begin{document} 

\maketitle
\flushbottom

\section{Introduction} \label{sec:intro}

Precision Electroweak (PEW) analysis is a powerful and well-established tool that tests the Standard Model (SM) and constrains theories of new physics (see e.g.~\cite{Freitas:2020kcn, Erler:2019hds, Wells:2005vk, Matchev:2004yw}).  The Standard Model has shown remarkable success at fitting observations  \cite{Flacher:2008zq,Baak:2012kk, Baak:2011ze, Baak:2014ora, Haller:2018nnx, Fan:2014vta}.  However, recent measurements including the $W$-boson mass by the CDF collaboration \cite{CDF:2022hxs} and the anomalous magnetic moment of the muon ($g-2$) \cite{Muong-2:2006rrc,Muong-2:2021ojo, Muong-2:2023cdq} have raised questions about the completeness of the Standard Model at the electroweak scale. In addition, the measurement of the forward-backward asymmetry in $b$ quarks at the $Z$ pole has historically been in  tension with the other PEW measurements \cite{ALEPH:2005ab, Baak:2011ze, Baak:2014ora, Haller:2018nnx}. 

These anomalous observations, along with the prospect of continued improvement of these and other PEW measurements motivate continued critical analysis of the implications of the PEW fit.  Here we briefly review each of these anomalous observations.

The recent measurement of the mass of the $W$ boson from the CDF collaboration \cite{CDF:2022hxs} is of comparable accuracy to the world average of all other measurements.  The measurement is puzzling because it is discrepant both with other measurements (including a recent one from the ATLAS collaboration \cite{ATLAS:2023fsi}) and with the Standard Model expectation as determined from fits to other precision electroweak observations. 

The anomalous magnetic moment of the muon has been measured with high precision and is incompatible with the SM expectation \cite{Muong-2:2006rrc, Muong-2:2021ojo, Muong-2:2023cdq, Aoyama:2020ynm, Keshavarzi:2021eqa}.  Even when the anomalous magnetic moment is not directly included in PEW fits, it still effects the fit indirectly because both the PEW fit and the muon magnetic moment require input from the hadronic vacuum polarization (HVP) diagram. For muon $g-2$, the error on this quantity represents the dominant contribution to the error budget of the Standard model prediction.  Intriguingly, there has been tension in the determination of this quantity using data from $\sigma(e^{+} e^{-} \rightarrow \; \rm{hadrons})$ on the one hand \cite{Aoyama:2020ynm, Keshavarzi:2019abf, Davier:2019can}, and a direct calculation using lattice QCD on the other \cite{Wittig:2023pcl, Borsanyi:2020mff}. In the case of the PEW fit, the HVP diagram allows (along with perturbative QED/QCD running) a determination of $\alpha(M_Z^2)$ from low energy measurements of the fine-structure constant.  The measurement of $\alpha(M_Z^2)$ is in turn used in the PEW fitting process. Modification of this vacuum polarization contribution can in principle bring the measurement of muon $g-2$ into agreement with the SM theoretical expectation, but at the cost of introducing tensions in the PEW fit and/or with the measurements of $e^{+} e^{-}$ to hadrons \cite{Passera:2008jk, Crivellin:2020zul, Keshavarzi:2020bfy, deRafael:2020uif, Malaescu:2020zuc}.

Finally, the forward-backward asymmetry to $b$-quarks, $A^{b,0}_{\text{FB}}$, as measured by LEP \cite{ALEPH:2005ab} has shown a consistent $\sim2.5\sigma$ tension with the Standard Model fit \cite{ALEPH:2005ab, Flacher:2008zq, Baak:2011ze, Baak:2014ora, Haller:2018nnx}.  That is, the effective weak mixing angle as determined by $A^{b,0}_{\text{FB}}$ and from other asymmetry parameters are not in agreement \cite{ALEPH:2005ab, Moortgat-Pick:2015lbx}.
It is not clear whether this is due to some unknown systematic effect, new physics that preferentially couples to heavy flavor, or just a statistical fluctuation. Future experiments have the potential to test each of these hypotheses; for some work on how experiments can probe the $Zb\Bar{b}$ coupling see Refs.~\cite{Yan:2023ccj, Dong:2022ayy, Li:2021uww, Yan:2021htf, Yan:2021veo}. For some experiments that probe the electroweak mixing angle see e.g., Refs.~\cite{Becker:2018ggl, MOLLER:2014iki, AbdulKhalek:2022hcn}.  For a more extensive discussion of current and future weak mixing angle measurements see Refs. \cite{ParticleDataGroup:2022pth, Davoudiasl:2023cnc}.

What to make of all these tensions remains obscure.  However, how we treat them has implications for the precision electroweak fit, and consequently, for constraints on physics beyond the Standard Model. In this work, we analyze each of these tensions and study their implications  within the framework of the oblique parameters \cite{Kennedy:1988sn, Peskin:1991sw, Peskin:1990zt}.

Next, we discuss the implications of these results for a new $U(1)_D$ symmetry whose corresponding gauge boson kinetically mixes with SM hypercharge \cite{Holdom:1985ag}.  
The kinetic mixing provides a renormalizable portal coupling from the SM to a potentially larger secluded sector. A new $U(1)$ represents a minimal extension to the Standard Model, and extra $U(1)$ symmetries are a generic expectation of many models of high-scale physics (see e.g., Refs.~\cite{Langacker:2008yv, Dienes:1996zr}). The gauge boson associated with the new $U(1)_D$ symmetry is often called a dark photon, dark $Z$, or $Z^{\prime}$.  We use these terms interchangeably in this work.  

This work builds upon the work of Refs.~\cite{Curtin:2014cca, Hook:2010tw}.  We study the combined impact of the PEW tensions on the precision electroweak fit and the dark photon parameter space.  We also discuss the effects of the existence of dark decay channels on direct searches for dark photons in  dilepton channels.  For recent work on PEW fits in light of the new CDF $W$ mass measurement as well as applications to the dark photon model see Refs.~\cite{Strumia:2022qkt, Asadi:2022xiy, Fan:2022yly, Gu:2022htv, Lu:2022bgw, deBlas:2022hdk, Zhang:2022nnh, Zeng:2022lkk, Thomas:2022gib, Cheng:2022aau, Cai:2022cti, Alguero:2022est, Du:2022fqv, Loizos:2023xbj}. For recent work that takes a different approach to reconciling the measurement of the anomalous magnetic moment of the muon using a dark photon, see \cite{Darme:2022yal, Coyle:2023nmi}.  

This work is organized as follows.  In Section \ref{sec:PEWAnalysis} we review the PEW fit and the oblique parameters.  In Section \ref{sec:KM} we discuss kinetic mixing and the dark photon.  Section \ref{sec:results} contains our main results.  We conclude in Section \ref{sec:conclusion}.  In an Appendix we comment on how and when oblique PEW constraints are insufficient to describe the effects of the dark photon.    

\section{Precision Electroweak Analysis}\label{sec:PEWAnalysis}

In this Section, we describe our precision electroweak fitting methodology.  We largely follow the approaches of Refs.~\cite{Curtin:2014cca, Haller:2018nnx}.  This section also provides a brief review of the oblique (Peskin-Takeuchi) parameters.

\subsection{Data and SM Predictions} \label{sec:SMpred}

Table \ref{tab:measurements} summarizes measured PEW observables. Many groups have performed fits to these precision measurements, notably the GFitter Group \cite{Flacher:2008zq, Baak:2011ze, Baak:2014ora, Haller:2018nnx}.  In the context of the dark photon model, fits were performed in Refs.~\cite{Hook:2010tw, Curtin:2014cca}.  We use the same set of observables as Ref.~\cite{Curtin:2014cca}. 

The PEW observables are measured sufficiently precisely that multi-loop SM computations are required for  comparison with experiment.  Parameterization formulae exist for these computations that highlight the dependence of the results on the input parameters.  These input parameters are the fine structure constant at zero energy $\alpha(q^2=0)$, the Fermi constant $G_F$, the mass of the $Z$ boson $M_Z$, the Higgs boson mass $M_h$, the top mass $M_t$, the strong coupling constant $\alpha_s (M_Z^2)$, and the hadronic contribution to the running of the fine structure constant $\Delta\alpha^{(5)}_{\text{had}}(M_Z^2)$.  For the relationship between $\alpha$ and  $\Delta\alpha^{(5)}_{\text{had}}$ see, e.g., Ref.~\cite{Erler:2019hds}. In practice, $\alpha(q^2=0)$ and $G_F$ are measured so precisely that they are effectively constants in the fit.

\begin{center}
\begin{table}
    \centering
    \begin{tabular}{c|c|c}
    Parameter & Measurement & Reference(s) \\
    \hline
    $\alpha^{-1}(q^2 = 0)$ & $137.035999180 \pm 0.000000010$ & \cite{ParticleDataGroup:2022pth} \\
    $G_F$ & $(1.166 378 8 \pm 0.000 000 6)\times 10^{-5}$ \GeV$^{-2}$ & \cite{ParticleDataGroup:2022pth} \\ 
    \hline\hline
    $M_t$ & $172.47 \pm 0.46$ \GeV & \cite{Haller:2018nnx, Haller:2022eyb} \\
    $M_h$ & $125.25 \pm 0.17$ \GeV & \cite{ParticleDataGroup:2022pth} \\
    $\alpha_s (M_Z^2)$ & --- & ---\\
    $\Delta\alpha^{(5)}_{\text{had}} (M_Z^2)$ &  $0.02768 \pm 0.00007$ & \cite{ParticleDataGroup:2022pth} \\
    $M_Z$  &  $91.1875 \pm 0.0021$ \GeV & \cite{Janot:2019oyi} \\
    \hline\hline
    $\Gamma_Z$ & $2.4955 \pm 0.0023$ \GeV & \cite{Janot:2019oyi}\\
    $\sigma^0_{\text{had}}$ &  $41.4802 \pm 0.0325$ nb & \cite{Janot:2019oyi}\\
    $R^0_{\ell}$ & $20.7666 \pm 0.0247$ & \cite{Janot:2019oyi} \\
    $A^{0,\ell}_{\text{FB}}$ & $0.0171 \pm 0.0010$ & \cite{Janot:2019oyi}\\
    $A_{\ell}$ & $0.1499 \pm 0.0018$ & \cite{Haller:2018nnx}\\
    $M_W$ [PDG 2022] & $80.377 \pm 0.012$ \GeV & \cite{ParticleDataGroup:2022pth}\\
    $M_W$ [2023 Combo] & $80.3946 \pm 0.0115$ \GeV & \cite{Amoroso:2023pey}\\
    $\Gamma_W$ & $2.085 \pm 0.042$ \GeV & \cite{ParticleDataGroup:2022pth} \\
    \hline
    $R^0_b$ & $0.21629 \pm 0.00066$ & \cite{ALEPH:2005ab, Haller:2018nnx}\\
    $R^0_c$ & $0.1721 \pm 0.0030$ & \cite{ALEPH:2005ab, Haller:2018nnx}\\
    $A^{0,b}_{\text{FB}}$ & $0.0992 \pm 0.0016$ & \cite{ALEPH:2005ab}\\
    $A^{0,c}_{\text{FB}}$ & $0.0707 \pm 0.0035$ & \cite{ALEPH:2005ab}\\
    $A_b$ & $0.923 \pm 0.020$ & \cite{ALEPH:2005ab, ParticleDataGroup:2022pth}\\
    $A_c$ & $0.670 \pm 0.027$ & \cite{ALEPH:2005ab, ParticleDataGroup:2022pth} 
    \end{tabular}
    \caption{Summary table of the measured values of the observables included in our fits.  The first two, $\alpha(q^2 = 0)$ and $G_F$, are measured to much higher precision then the others and are fixed in our fits.  The observables between the double lines, $M_Z, M_h, M_t,  \alpha_s (M_Z^2), \Delta\alpha^{(5)}_{\text{had}} (M_Z^2)$, are free parameters in the fit.  Following Refs.~\cite{Haller:2018nnx, Curtin:2014cca}, we do not use the measured value of $\alpha_s$ in the fit.  The last 6 observables contain end-state $b$- and $c$-quarks. We call these the heavy flavor (HF) observables.  Following Ref.~\cite{Curtin:2014cca} we exclude $\text{sin}^2 \theta^l_{\text{eff}}(Q_{\text{FB}})$ from the fit.  The 2022 PDG $W$ mass average \cite{ParticleDataGroup:2022pth} does not include the latest measurements from CDF \cite{CDF:2022hxs} or ATLAS \cite{ATLAS:2023fsi}.  The 2023 Combo $W$ mass average \cite{Amoroso:2023pey}  includes the latest measurements. }
    \label{tab:measurements}
\end{table}
\end{center}

For the $Z$ boson width $\Gamma_z$, total peak hadronic cross section $\sigma_{\text{had}}^0$, and the ratios $R_{\ell} \equiv \Gamma_{\text{had}}/\Gamma_{\ell}$, $R_b \equiv \Gamma_b/\Gamma_{\text{had}}$, and $R_c \equiv \Gamma_c/\Gamma_{\text{had}}$ we use the parameterization formulae found in Ref.~\cite{Dubovyk:2018rlg}.  For the mass of the $W$ boson, the parameterization formula in Ref.~\cite{Awramik:2003rn} is used. For the $W$ boson width $\Gamma_W$ we use the parameterizations of Ref.~\cite{Cho:2011rk}.  To get the forward-backward asymmetry parameters from the effective weak mixing angles we follow the standard prescription as laid out in, e.g., Ref.~\cite{Flacher:2008zq}.  We require expressions for the effective weak mixing angles for leptons, $c$ quarks, and  $b$ quarks. 
For $\text{sin}^2\theta^l_{\text{eff}}$ and $\text{sin}^2\theta^c_{\text{eff}}$ we use the parameterization formulae from Ref.~\cite{Awramik:2006uz}.  For $\text{sin}^2\theta^b_{\text{eff}}$ we utilize the formula from Ref.~\cite{Dubovyk:2016aqv}.\footnote{In this formula, we have corrected the factor of 10 typographic error in the contribution that depends on  $\Delta\alpha$.} 

\subsection{Oblique Parameters}
\label{sec:oblique}
 
There are many ways to parameterize the effects of new physics on physical observables. We focus on the oblique (Peskin-Takeuchi) parameters \cite{Kennedy:1988sn, Peskin:1991sw, Peskin:1990zt} that capture modifications to the gauge boson self-energies. The oblique corrections are defined by 
\begin{subequations}
\begin{align}
    S &\equiv \frac{4c_w^2s_w^2}{\alpha(M_Z^2)}\left(\frac{\Pi_{ZZ}(M_Z^2)-\Pi_{ZZ}(0)}{M_Z^2}  - \frac{c_w^2 - s_w^2}{c_w s_w}\frac{\Pi_{Z\gamma}(M_Z^2)}{M_Z^2} - \frac{\Pi_{\gamma\gamma}(M_Z^2)}{M_Z^2}\right)\\
    T &\equiv \frac{1}{\alpha(M_Z^2)}\left(\frac{\Pi_{WW}(0)}{M_W^2} - \frac{\Pi_{ZZ}(0)}{M_Z^2}\right)\\
    U &\equiv \frac{4s_w^2}{\alpha(M_Z^2)}\left(\frac{\Pi_{WW}(M_W^2)-\Pi_{WW}(0)}{M_W^2}  - \frac{c_w}{s_w}\frac{\Pi_{Z\gamma}(M_Z^2)}{M_Z^2} - \frac{\Pi_{\gamma\gamma}(M_Z^2)}{M_Z^2}\right) - S.
\end{align}
\end{subequations}
Here the $\Pi(Q)$'s are vacuum polarizations evaluated at energy scale $Q$, and $s_w$ and $c_w$ are the sine and the cosine of the weak mixing angle.  See Refs.~\cite{Freitas:2020kcn,Wells:2005vk,Matchev:2004yw} for pedagogical reviews. Only new physics contributions to the vacuum polarizations are included in the above expression.  The SM predictions are subtracted off, so $S=T=U=0$ corresponds to the Standard Model.

The oblique corrections can also be understood in terms of modifications to the gauge boson kinetic and mass terms in the Lagrangian \cite{Burgess:1993vc}.   Suppose one adds the following terms to the SM Lagrangian \cite{Burgess:1993vc}
\begin{equation}
\begin{aligned}
\hat{\mathcal{L}}_{\mathrm{new}}= & -\frac{A}{4} \hat{F}_{\mu \nu} \hat{F}^{\mu \nu}-\frac{B}{2} \hat{W}_{\mu \nu}^{\dagger} \hat{W}^{\mu \nu}-\frac{C}{4} \hat{Z}_{\mu \nu} \hat{Z}^{\mu \nu}+\frac{G}{2} \hat{F}_{\mu \nu} \hat{Z}^{\mu \nu} \\
& -w \tilde{m}_W^2 \hat{W}_\mu^{\dagger} \hat{W}^\mu-\frac{z}{2} \tilde{m}_z^2 \hat{Z}_\mu \hat{Z}^\mu,
\end{aligned}
\end{equation}
where the hats indicate that in the presence of these new terms, the kinetic terms are not canonically normalized.  Here, $A$, $B$, $C$, $G$, $w$, and $z$ are small constants, only three of which are independent due to the freedom to make field redefinitions.   The oblique parameters are one particular choice of the three independent degrees of freedom given by \cite{Burgess:1993vc}:

\begin{equation}
\begin{aligned}
& \alpha S=4 s_w^2 c_w^2\left(A-C-\frac{c_w^2-s_w^2}{c_w s_w} G\right), \\
& \alpha T=w-z, \\
& \alpha U=4 s_w^4\left(A-\frac{1}{s_w^2} B+\frac{c_w^2}{s_w^2} C-2 \frac{c_w}{s_w} G\right) .
\end{aligned}
\end{equation}

When one canonically normalizes the kinetic terms and rewrites all the SM parameters in terms of $e$, $s_w$ and $M_Z$, the oblique parameters lead to the following modifications to the couplings of the $Z$ boson \cite{Burgess:1993vc, Babu:1997st}
\begin{equation}\label{eq:ZcoupsOblique}
\mathcal{L}_{\text{$Z$ Couplings}} = -\frac{e}{2s_w c_w}\left(1+\frac{\alpha T}{2}\right)\sum_i\Bar{\psi}_i\gamma^{\mu}((T_3^i - 2Qs^2_{\star}) - T_3^i\gamma^5)\psi_iZ_{\mu},
\end{equation}
where the quantity $s_{\star}$ is defined by
\begin{equation}\label{eq:ZcoupsOblique2}
    s_{\star}^2 = s_w^2 + \frac{\alpha}{c_w^2 - s_w^2}\left(\frac{S}{4} - c_w^2 s_w^2 T \right).
\end{equation}
In addition, the oblique parameters lead to a modification to the relationship between the $W$ and $Z$ masses.  Explicitly, $M_W$ is now given by

\begin{equation} \label{eq:WMassExpr}
    M_W = M_{W, SM}\left[1 - \frac{\alpha(M^2_Z)}{4(c_w^2 - s_w^2)}\left(S - 2c_w^2T - \frac{c_w^2 - s_w^2}{2s_w^2}U\right)\right].
\end{equation}

In many models of new physics, $U \ll S,T$.  One way to understand why this is so, is to connect the oblique parameters to higher-dimensional SM effective field theory operators that contain the gauge bosons and the Higgs field.  Explicitly S, T, and U are proportional to the Wilson coefficients of \cite{Grzadkowski:2010es, Han:2004az, Han:2008es}
\begin{subequations}
    \begin{align}
        S &\leftrightarrow (h^{\dagger}\sigma^ah)W^a_{\mu \nu}B^{\mu \nu}\\
        T &\leftrightarrow (h^{\dagger}D^{\mu}h)(D_{\mu}h^{\dagger}h)\\
        U &\leftrightarrow (h^{\dagger}W^{a\mu\nu}h)(h^{\dagger}W^a_{\mu\nu}h).
    \end{align}
\end{subequations}
Notice that $U$ is related to a dimension-8 operator while the operators associated with $S$ and $T$ are dimension-6. This connection to higher dimensional operators is useful to understand why $U$ is small compared to $S$ and $T$ in many models of new physics.  However, the effective field theory language is less useful when there is not a clear separation of scales.  In the dark photon model, it is only the case that $U\ll S, T$ when the dark photon mass is much greater than the SM $Z$ boson.  In Section \ref{sec:kmST} we will derive $S$, $T$, and $U$ in the dark photon model.  In the first part of our analysis, we will set $U = 0$ to allow for applicability to a wide range of new physics models.  Later, when we specialize to the case of the dark photon, we will allow for $U\neq 0$.

The PEW observables can be written as the SM prediction plus contributions due to $S$, $T$, and $U$ ($\mathcal{O} = \mathcal{O}_{SM} +  \#S +  \#T + \#U$).  These expressions are summarized in Appendix A of Ref.~\cite{Ciuchini:2013pca} (See also Refs.~\cite{Peskin:1991sw, Burgess:1993mg}).  For an explicit example of one of these expressions see Eq.~(\ref{eq:WMassExpr}) above.  We fix the value of the weak mixing angle appearing in these expressions via
\begin{equation}\label{eq:swdef}
    s_w^2c_w^2 = \frac{\pi \alpha(M_Z^2)}{\sqrt{2} G_F M_Z^2}.
\end{equation}
Expressions like Eq.~(\ref{eq:WMassExpr}) give observables as a function of the oblique parameters and a SM prediction.  For the SM prediction, we use the parameterization formulae discussed above in Section ~\ref{sec:SMpred}.  The end result is a theoretical prediction for each observable that is a function both of the SM parameters and $S$, $T$, and $U$.

\subsection{Fitting Procedure}\label{sec:methodology}

To perform the fit, we construct a Chi-Square statistic,
\begin{equation}
    \chi^2(\Vec{\theta} = M_Z, M_h, M_t, \alpha_s, \Delta\alpha^{(5)}_{\text{had}}, S, T, U) = (\Vec{y} - \Vec{\mu}(\Vec{\theta}))^{T}V^{-1}(\Vec{y} - \Vec{\mu}(\Vec{\theta}))
\end{equation}
where $\Vec{y}$ is a vector of the measurements from Table \ref{tab:measurements}, $\Vec{\mu}(\Vec{\theta})$ is a vector of the theoretical predictions in terms of the free parameters in the fit that utilizes expressions as described in the previous section, and $V$ is the covariance matrix of the observables.   The covariance matrix is given by \cite{Curtin:2014cca}
\begin{equation}
    V = \Sigma\rho\Sigma
\end{equation}
where $\Sigma = \text{diag}(\sigma_1, \sigma_2, ...)$ is a matrix with the experimental uncertainties of the measurements along the diagonal and $\rho$ is the experimental correlation matrix as reported by Refs.~\cite{ALEPH:2005ab, Janot:2019oyi}.  We neglect theoretical uncertainties in our fit. Assuming that the observables are Gaussian one can interpret this chi-square function as $\chi^2 = -2 \ln L \, +$  constant where $L$ is the likelihood.

When putting bounds on the oblique parameters,  $M_Z, M_h, M_t, \alpha_s, \Delta\alpha^{(5)}_{\text{had}}$ are nuisance parameters.\footnote{As noted in Table \ref{tab:measurements}, the measurement of $\alpha_s$ is not used in the construction of the chi-square function.  However, the chi-square function depends on $\alpha_s$ parametrically via the parameterization formulae discussed in Section  \ref{sec:SMpred}.}.  To eliminate the dependence on these nuisance parameters, we use a profiling approach.  In particular, we define $\hat{\chi}^2(S, T, U) = \chi^2(\hat{M}_Z, \hat{M}_h, \hat{M}_t, \hat{\alpha}_s, \hat{\Delta\alpha}^{(5)}_{\text{had}}, S, T, U)$ where the hat indicates that for each $S$, $T$, and $U$ we pick the values of the nuisance parameters that minimize the chi-square value.  For certain parts of the analysis we will set $U=0$.  After fixing $U = 0$, we identify the $2\sigma$ (95.45\%) confidence region for $S$ and $T$ by requiring $\Delta\hat{\chi}^2(S, T) = \hat{\chi}^2(S, T) - \chi^2_{\text{min}} = 6.18$ where $\chi^2_{\text{min}}$ is the global minimum.  The preferred regions in the $S$-$T$ plane for different data combinations will be shown in Section \ref{sec:results}.

\section{Kinetic Mixing and the Dark Photon}\label{sec:KM}

The hypercharge field-strength tensor is dimension 2 and invariant under the $SU(3)\times SU(2)_L \times U(1)_Y$ gauge symmetry of the Standard Model.  In the presence of a new $U(1)_D$ gauge symmetry (under which the SM fields are singlets), the hypercharge field strength can kinetically mix with the $U(1)_D$ field strength.  This dimension-4 renormalizable ``vector portal'' is a well-studied extension to the Standard Model \cite{Holdom:1985ag, Curtin:2014cca, Fabbrichesi:2020wbt, Gopalakrishna:2008dv, Babu:1997st}.  
  
There are two common routes used to give the dark photon a mass.  The first is known as the  Stueckelberg mechanism \cite{Stueckelberg:1938hvi, Feldman:2007wj, Ruegg:2003ps, Kors:2004dx}, wherein one introduces a new scalar field $\Phi_S$ that transforms appropriately under the $U(1)_D$ symmetry such that a term like $\mathcal{L} \supset (\partial_{\mu}\Phi_S + M Z_{\mu})^2$ is gauge invariant.  This term  gives a mass to the gauge boson and no propagating scalar degree of freedom. The second is to spontaneously break the $U(1)_D$ symmetry by introducing a new, ``dark Higgs'' field $\Phi_D$ charged under $U(1)_D$ and arranging for it to take on a vacuum expectation value.  This dark Higgs field can have a portal coupling to the SM Higgs field $\Phi_{SM}$ via a term in the Lagrangian that looks like
\begin{equation}
    \mathcal{L} \supset \kappa |\Phi_D|^2|\Phi_{SM}|^2.
\end{equation}
The addition of this additional field and portal coupling to the SM Higgs field can have interesting phenomenological consequences (see e.g. \cite{Gopalakrishna:2008dv, Curtin:2014cca}).  For the purposes of this work, we assume $\kappa$ is small enough that these are negligible. The smallness of the Higgs portal coupling is radiatively stable and is technically natural. Also, the smallness of the tree-level quartic may be understood by (not necessarily low-scale) supersymmetry that constrains quartic couplings, or if the dark gauge boson gets its mass via dimensional transmutation of an additional non-Abelian dark group (i.e., a technicolor like mechanism)~\cite{Harigaya:2016rwr,Contino:2020god}. In the latter case the portal coupling would be suppressed by powers of the dimensional transmutation scale in the dark sector $\Lambda_{Dark}$ to the cutoff scale.  Our analysis also applies if the mass of the new gauge boson was generated via the Stueckelberg mechanism. 

Following the notations and conventions of Ref.~\cite{Davoudiasl:2012ag, Curtin:2014cca}, we now present basic information about the dark photon model.  We write the relevant parts of the Lagrangian as
\begin{equation}\label{eq:km_lag}
    \mathcal{L} \supset -\frac{1}{4}\hat{B}_{\mu \nu}\hat{B}^{\mu \nu}  -\frac{1}{4}\hat{Z}_{D \mu \nu}\hat{Z}_D^{\mu \nu} + \frac{\epsilon}{2 c_w}\hat{Z}_{D \mu \nu}\hat{B}^{\mu \nu} + \frac{1}{2}M^2_{D,0}\hat{Z}_{D \mu}\hat{Z}_D^{\mu},
\end{equation}
where $Z_{D \mu}$ is the dark photon, $B_{\mu}$ is the hypercharge boson, and $c_w = \text{cos}\theta_w$. Note $\epsilon$ is constrained to be  small parameter to avoid experimental constraints.  If it arises at one loop by integrating out fields charged under both $U(1)_Y$ and $U(1)_D$ \cite{Holdom:1985ag}, it might be expected to be of the order $10^{-3}$ (See Ref.~\cite{Gherghetta:2019coi} for further discussion of the magnitude of $\epsilon$).  The hats over the fields indicate that they are not canonically normalized and a subscript $0$ on the mass parameter is used to indicate that the fields are not in the mass eigenbasis.  

To get the Lagrangian of Eq.~(\ref{eq:km_lag}) into canonical form, one must first perform a field redefinition to normalize the kinetic terms.  The transformation is of the form
\begin{equation}\label{eq:kindiagonalization}
    \begin{pmatrix}
    Z_{D,0}\\
    B
    \end{pmatrix}
    =\begin{pmatrix}
        \sqrt{1 - \epsilon^2/c_w^2} & 0\\
        -\epsilon/c_w & 1
    \end{pmatrix}
    \begin{pmatrix}
        \hat{Z}_{D,0}\\
        \hat{B}
    \end{pmatrix}
\end{equation}
where again, the subscript 0 indicates that this field is not yet a mass eigenstate. 

The next step is to diagonalize the gauge boson mass terms. It is convenient to define the following quantities
\begin{equation}
    \eta \equiv \frac{\epsilon/c_w}{\sqrt{1-\frac{\epsilon^2}{c_w^2}}}, \qquad \qquad \delta^2 \equiv \frac{M^2_{D,0}}{M^2_{Z,0}}.
\end{equation}
We write the necessary rotation as
\begin{equation}\label{eq:massdiagonalization}
    \begin{pmatrix}
    Z\\
    Z_D
    \end{pmatrix}
    =\begin{pmatrix}
        \text{cos}\theta_{\alpha} & \text{sin}\theta_{\alpha}\\
        -\text{sin}\theta_{\alpha} & \text{cos}\theta_{\alpha}
    \end{pmatrix}
    \begin{pmatrix}
        Z_{0}\\
        Z_{D,0}
    \end{pmatrix}
\end{equation}
where the mass mixing angle $\theta_{\alpha}$ can be written \cite{Gopalakrishna:2008dv, Curtin:2014cca, Babu:1997st}
\begin{equation}
\label{eq:massrot}
    \text{tan}2\theta_{\alpha} = \frac{-2 s_w \eta}{1- \delta^2 - s_w^2 \eta^2}.
\end{equation}
The mass eigenvalues are \cite{Curtin:2014cca}
\begin{equation}
    M^2_{Z, Z_D} = \frac{M_{Z, 0}^2}{2}\left(1 + \delta^2 + s_w^2\eta^2 \pm \text{Sign}(1-\delta^2)\sqrt{(1+\delta^2 + s_w^2\eta^2)^2 - 4\delta^2}\right).
\end{equation}
As long as the masses are not degenerate, we can write
\begin{equation}
    r^2 \equiv \frac{M^2_{Z_D}}{M^2_Z} = \frac{M^2_{D,0}}{M^2_{Z,0}} + \mathcal{O}(\epsilon^2)
\end{equation}
where we have defined $r$ to be the mass ratio of the dark photon mass to the SM $Z$ mass.

\subsection{Kinetic Mixing and the Oblique Parameters} \label{sec:kmST}

At the $Z$-peak, the dominant effect of the dark photon is to modify the $Z$-boson's mass and couplings.  This effect is captured by the oblique parameters.  We will comment on exceptions at the end of this section.

To find the oblique parameters in terms of the kinetic mixing parameters, one can start with the kinetic mixing Lagrangian of Eq.~(\ref{eq:km_lag}), canonically normalize the kinetic terms, fix the physical constants to their SM values, and then match to the general form of the mass terms and interactions in the presence of oblique corrections.  The result, up to quadratic order in the kinetic mixing parameter $\epsilon$, is \cite{Burgess:1993vc, Davoudiasl:2023cnc, Holdom:1990xp}
\begin{subequations}\label{eq:STUinKM}
    \begin{align}
        S &= \frac{4 s_w^2 (c_w^2 - r^2)}{\alpha (M_Z^2)}\left(\frac{\epsilon}{1-r^2}\right)^2\\
        T &= \frac{-s_w^2 r^2}{c_w^2\alpha(M_Z^2)}\left(\frac{\epsilon}{1-r^2}\right)^2\\
        U &= \frac{4s_w^4}{\alpha(M_Z^2)}\left(\frac{\epsilon}{1-r^2}\right)^2,
    \end{align}
\end{subequations}
where $r = M_{Z_D}/M_Z$ as defined above.

When the kinetic mixing parameter vanishes, so do $S$, $T$ and $U$, i.e. $\epsilon = 0 \implies S = T = U =0$.  In the limit of a heavy dark photon, $U\ll S,T$ and the ratio $T/S$ is given by 
\begin{equation} \label{eq:STRatio}
    \frac{T}{S} = \frac{-r^2}{4c_w^2(c_w^2 - r^2)}  \qquad \xrightarrow{r \gg 1} \qquad \frac{1}{4 c_w^2}.
\end{equation}
For a derivation of $S$, $T$, and $U$ valid in the limit of a heavy dark photon see Ref. \cite{Babu:1997st}.

The dark photon does not affect the leading order expressions for non-$Z$-pole observables in terms of the model parameters.  However, the presence of the dark photon modifies the expressions for the $Z$-pole observables in terms of these parameters. So, once these observables are fixed to their physical values, the inferred values for the model parameters can shift.  This, in turn, can shift the non-$Z$-pole observables relative to the SM expectation.  As an example, consider the mass of the $W$ boson.  Upon fixing $G_F$, $\alpha$, and $M_Z$ to take on their measured values and defining the weak mixing angle $s_w$ via Eq.~(\ref{eq:swdef}),  we can write the mass of the $W$ boson in terms of the kinetic mixing model parameters as 
\begin{equation}
    M_W = M_{W, SM}\left(1 - \frac{s_w^2 \epsilon^2}{2(c_w^2 - s_w^2) (1-r^2)}\right),
\end{equation}
where $M_{W, SM}$ is the SM expectation for the mass of the $W$ boson.  The above expression is valid so long as $\epsilon$ is small and the mass ratio $r$ is not equal to 1.  Note this matches the result of plugging in the expressions for $S$, $T$, and $U$ given in Eqs.~(\ref{eq:STUinKM}) into Eq.~(\ref{eq:WMassExpr}).

To reinterpret the bounds on the oblique parameters in terms of the dark photon parameters we can use the  expressions for $S$, $T$, and $U$ in Eqs.~(\ref{eq:STUinKM}) in terms of  $M_{Z_D}$ and $\epsilon$ (or equivalently $\eta$).  Explicitly, for the 2$\sigma$
confidence region in the $M_{Z_D}$-$\eta$ plane we find where $\Delta\hat{\chi}^2(M_{Z_D}, \eta) = \hat{\chi}^2(S(M_{Z_D}, \eta), T(M_{Z_D}, \eta), U(M_{Z_D}, \eta)) - \chi^2_{\text{min}} = 6.18$.  This confidence region is shown in Section \ref{sec:results}.  The mass region of greatest interest is $M_Z \lesssim M_{Z_D} \lesssim 400 \; \GeV$.

For $Z_{D}$ masses very close to the $Z$-boson mass (within a few GeV), there will be non-oblique corrections to precision electroweak observables, even on the $Z$-pole.  Moreover, extracting SM parameters in the presence of such a nearly degenerate  $Z_{D}$ can prove challenging.  For more discussion, see Appendix \ref{sec:nonoblique}.  For a recent work on a nearly-degenerate dark photon and $Z$ boson system see Ref. \cite{Qiu:2023zfr}.  In the main text, we will assume that such subtleties are absent.

\subsection{Collider Bounds} \label{sec:colliderbounds}

Colliders set important constraints on the parameters of the dark photon model.  Direct searches at hadron colliders, working under the assumption that there are no decays of $Z_{D}$ to a dark sector,  typically require $\epsilon \lesssim \text{few}\times 10^{-3}$ over the region of parameter space $M_Z \lesssim M_{Z_D} \lesssim 400 \; \GeV$.
LHC constraints can be found in Refs.~\cite{San:2022uud, CMS:2019buh, ATLAS:2019erb, CMS:2021ctt}.\footnote{The small numerical difference between the bounds in Ref.~\cite{San:2022uud} and Ref.~\cite{CMS:2019buh} in the overlapping areas of parameter space can be attributed to differences in partonic luminosity functions used to translate between partonic and hadronic cross sections.  We use the bounds from \cite{San:2022uud} for consistency across the whole mass range.} (For a summary of collider constraints at lower dark photon masses see e.g. Ref.~\cite{Graham:2021ggy}; For discussion of future collider constraints at the HL-LHC and at a future muon collider see Ref. \cite{Hosseini:2022urq}).  Since PEW constraints are typically about an order of magnitude less stringent (i.e. $\epsilon \lesssim \text{few}\times 10^{-2}$), conventional wisdom says that the PEW constraints are less important.  However, there are exceptions to this rule.

First, if the dark photon has a mass in the window $75 < M_{Z_D} < 110\ \GeV$ there are no LHC constraints due to a veto around the SM $Z$ pole in the relevant analysis \cite{CMS:2019buh}.  We will argue below in Section \ref{sec:results} that searches in this region are  motivated, despite the challenges associated with large backgrounds from the SM $Z$ boson.

Second, as mentioned above, collider bounds depend on the decays of the dark photon, while PEW bounds are decay agnostic.\footnote{Another decay-agnostic dark photon search method is deep inelastic scattering of electrons or positrons off protons.  In our dark photon mass range of interest, these bounds are currently weaker than the PEW bounds.  However, future deep inelastic scattering experiments have the potential to supersede the PEW bounds \cite{Kribs:2020vyk}. For other related work see e.g. Refs. \cite{Thomas:2021lub, Yan:2022npz, Thomas:2022qhj}}.  Collider bounds can be substantially relaxed if the dark photon has significant branching ratio to dark sector states. The production cross section of the dark photon goes like the kinetic mixing parameter squared, i.e., $\sigma(q\Bar{q}\rightarrow Z_D) \sim \epsilon^2$. 
Because all decay widths to SM particles  scale as $\epsilon^2$, if the dark photon can only decay to SM states then the branching ratio to muons Br$(Z_D \rightarrow \mu^+\mu^-) = \Gamma(Z_D \rightarrow \mu^+\mu^-)/\Gamma(Z_D \rightarrow SM)$ is independent of $\epsilon$, and the cross section times branching ratio will retain the $\epsilon^2$ scaling of the production cross section. On the other hand, if there are some dark sector states to which the dark photon decays such that $\Gamma(Z_D \rightarrow \text{dark}) \gg \Gamma(Z_D \rightarrow f\Bar{f})$ then we have 
\begin{equation}
\label{eq:epsfourth}
    \sigma \frac{\Gamma(Z_D \rightarrow \mu^+\mu^-)}{\Gamma_{Z_D, \text{tot}}} \approx  \sigma \frac{\Gamma(Z_D \rightarrow \mu^+\mu^-)}{\Gamma_{Z_D, \text{dark}}}\sim \epsilon^4.
\end{equation}
The original bounds, arising from searches for narrow dimuon resonances \cite{San:2022uud, CMS:2019buh, ATLAS:2019erb, CMS:2021ctt}, assume the dark photon exclusively decays to SM states.  Below, we recast these bounds as a function of the width to dark sector states, i.e., accounting for the modified scaling of Eq.~(\ref{eq:epsfourth}). 

Let $B$ denote the original bound on cross section times branching ratio obtained under the assumption that the $Z_{D}$ decays exclusively to SM final states.  Then
\begin{equation}\label{eq:borig}
    \sigma \times \text{Br} < B \quad \implies \quad \epsilon^2 < \frac{B}{\sigma/\epsilon^2}\frac{\Gamma(Z_D \rightarrow SM)}{\Gamma(Z_D \rightarrow \mu^+\mu^-)} \equiv b_{\rm orig},
\end{equation}
where we have defined $b_{\rm orig}$ as the original bound on $\epsilon^2$.
Now consider the case  where decays to dark-sector states dominate, $\Gamma(Z_D \rightarrow \text{dark}) \gg \Gamma(Z_D \rightarrow SM)$ so that $\Gamma(Z_D \rightarrow \text{dark}) \approx \Gamma_{Z_D, \text{tot}}$.
In this scenario, the same bound on cross section times branching ratio leads to a different bound on $\epsilon$
\begin{equation}\label{eq:bnew}
    \sigma \times \text{Br} < B \quad \implies \quad \epsilon^2 < \left(\frac{B}{\sigma/\epsilon^2}\frac{\Gamma_{Z_D, \text{tot}}}{\Gamma(Z_D \rightarrow \mu^+\mu^-)/\epsilon^2}\right)^{1/2} \equiv b_{\rm new},
\end{equation}
where we have defined $b_{\rm new}$ as the new bound on $\epsilon^2$.  Comparing Eqs.~(\ref{eq:bnew}) and ~(\ref{eq:borig}), we see that 
\begin{equation}
    b_{\rm new} = \left(b_{\rm orig}\times\frac{\Gamma_{Z_D, \text{tot}}}{\Gamma(Z_D \rightarrow SM)/\epsilon^2}\right)^{1/2}.
\end{equation}

The presence of light dark-sector states is a reasonable and  minimal extension to the dark-photon model.  It is tempting to assume that (one of) these states comprise the dark matter.  However, in this case dark matter direct-detection experiments can introduce additional constraints.  We briefly discuss this in Section \ref{sec:DarkMatter}. Even in the case that these light dark states are not cosmologically stable, there is still a constraint \cite{Hook:2010tw} arising from the measurement of the invisible width of the $Z$ boson \cite{ALEPH:2005ab, ParticleDataGroup:2022pth}.  Due to the small rotation necessary to diagonalize the vector boson mass terms,  see Eq.~(\ref{eq:massrot}), the SM-like $Z$ boson acquires a small coupling to these new dark sector states.  In the limit where the masses of the dark sector final states can be neglected, we can write \cite{Hook:2010tw}
\begin{equation}\label{eq:Zinvswid}
    \Gamma(Z \rightarrow \text{dark}) = \text{tan}^2\theta_{\alpha}\frac{M_Z}{M_{Z_D}}\Gamma(Z_D\rightarrow\text{dark}).
\end{equation}
In addition, in the presence of a dark photon, the form of the $Z$ boson neutrino coupling is modified leading to a modification in the prediction for the the $Z$ boson partial width to neutrinos. 
To find the bound due to the invisible width, we first compute the total invisible width in the context of the dark photon model $\Gamma(Z \rightarrow \text{Invs, tot}) = \Gamma(Z \rightarrow \text{dark}) + \Gamma(Z \rightarrow \nu\nu)$ where $\Gamma(Z \rightarrow \nu\nu)$ is modified to account for the shift in coupling shown in Eq.~(\ref{eq:ZcoupsOblique}).  Then, we construct a chi-square statistic by comparing this theoretical prediction to the measured value of the invisible width as reported in the Electroweak Model and Constraints on New Physics Review in Ref. \cite{ParticleDataGroup:2022pth}. The constraint from the invisible width of the $Z$ boson on the coupling of a dark photon to a dark sector is also examined in~\cite{Loizos:2023xbj}, although the impact of the coupling on the LHC constraint is not investigated.

\begin{figure}[t!]
    \centering
    \includegraphics[scale=1.0]{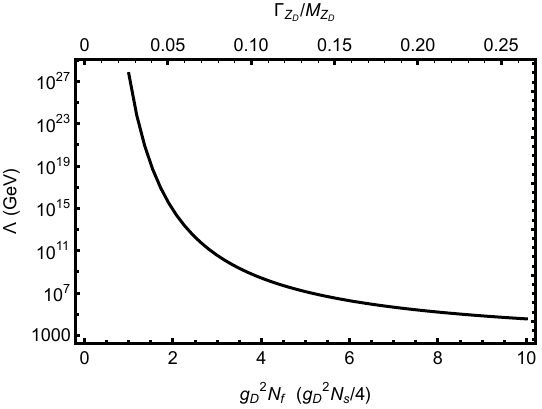}
    \caption{Location of the Landau pole  $\Lambda$ that results from adding $N_{f}$ new light Dirac fermions to which the $Z_{D}$ may decay.  On the $x$-axis, at bottom we show as a function of $g_{D}$ and $N_{f}$. At top, we show the corresponding fractional width $(\Gamma_{Z_D}/M_{Z_D})$ of the dark gauge boson.  As noted in the text, the result is readily interpreted in terms of complex scalars by making the replacement of the $x$-axis as: $g_D^2 N_f \rightarrow g_D^2 N_s/4$.}
    \label{fig:LandauPolePlot}
\end{figure}

One might ask what widths are reasonable to expect. Fig.~\ref{fig:LandauPolePlot} shows the Landau Pole of the $U(1)$ dark gauge interaction as a function of $g_D^2 N_f$, where $g_D$ is the $U(1)$ dark gauge coupling and $N_f$ is the number of Dirac fermions with a $U(1)$ dark charge of unity.  We identify the Landau pole by calculating the one-loop beta function for a $U(1)$ theory and then seeing where the coupling diverges.   We assume that the new degrees of freedom are fermionic and vectorlike.   At the top of the figure we show the corresponding fractional width $(\Gamma_{Z_D}/M_{Z_D})$.

Note that calculations  of the decay are done at tree-level, so results should be merely taken as illustrative, especially at large values of $g^2 N_{f}$. This shows that a fractional width of ${\mathcal O}$(1\%) of $M_{Z_{D}}$ is straightforward to achieve, and even larger widths may be achieved without running afoul of a Landau pole.  However, for widths that are too large, one might wonder whether our tree-level calculations of PEW effects resulting from the dark sector are accurate.  Moreover, the constraints from the invisible width of the $Z$ boson discussed above become stronger. Also note that the results for fermions can be easily translated to those for complex scalars.  Each complex scalar contributes $1/4$ as much as a Dirac fermion to both the width and the running, so that the $x$-axis can also be re-interpreted as $g_{D}^2 N_{S}/4$ as indicated. 

\subsection{Dark Matter}
\label{sec:DarkMatter}

As discussed above, introducing new dark states can substantially weaken limits that come from direct searches for the $Z_{D}$ at the LHC.  Here we comment briefly on the possibility that this dark state comprises the dark matter. For related ideas and studies, see \cite{Arkani-Hamed:2008hhe,Chun:2010ve,Evans:2017kti, Alves:2013tqa, Lebedev:2014bba, Arcadi:2013qia, Aboubrahim:2022qln, Cassel:2009pu}. 

A spin-independent coupling to nucleons mediated by the $Z/Z_D$ is strongly bounded by direct detection experiments, see e.g., \cite{LZ:2022ufs}.   In the case of fermionic dark matter, these strong bounds may be avoided in the presence of a small Majorana mass, which has the effect of splitting the would be Dirac dark matter into two Majorana states \cite{Tucker-Smith:2001myb}.  The spin-independent coupling is then off-diagonal between these two  states. Any splitting greater than 200 keV is sufficient to strongly kinematically suppress this scattering.  However, such a small splitting is kinematically irrelevant for decays of the $Z_{D}$, so the large decay width to dark states can be maintained while removing direct detection signals. A similar approach can be taken if the dark matter were a complex scalar.  In this case, the direct detection bounds can be evaded by splitting it into two real scalars \cite{Han:1997wn,Hall:1997ah}.

\section{Results}\label{sec:results}

In this Section, we present the results of our PEW fits. In addition to the standard set of 2022 PDG measurements \cite{ParticleDataGroup:2022pth} we explore the effects of modifying observables associated with $M_{W}$, $\Delta \alpha^{(5)}_{\text{had}}$, and $A_{FB}^{b}$ as discussed in the introduction. 

\begin{figure}
    \centering
    \includegraphics[scale=1.0]{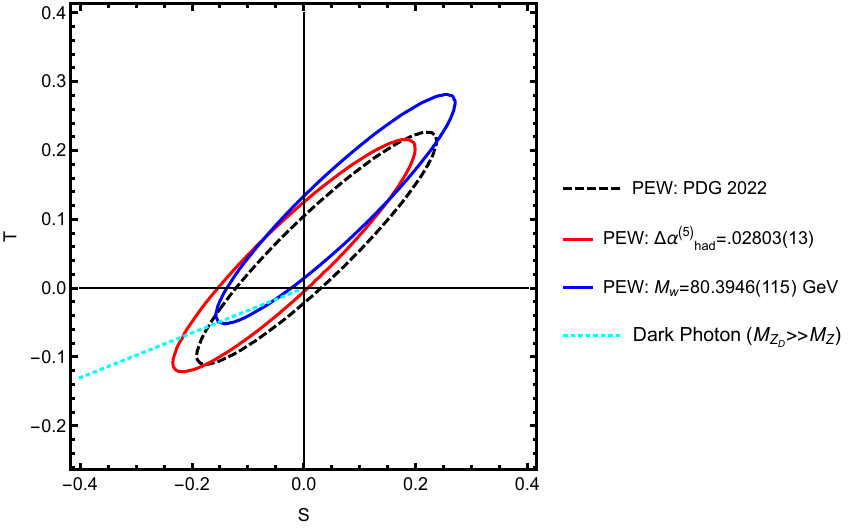}
    \caption{The 2$\sigma$ region in the $S$-$T$ plane (with $U$=0) for different data combinations using all the observables listed in Table \ref{tab:measurements}.  The black dashed contour shows the result using the measurements in the 2022 PDG \cite{ParticleDataGroup:2022pth}. The blue contour shows the result when the CDF measurement is included in the average for $M_W$ \cite{Amoroso:2023pey}.  The red line shows the result of shifting the hadronic contribution to the running of the fine structure constant \cite{Crivellin:2020zul}.  The dotted cyan line is the line traced out by kinetically mixed dark photon model in the limit $M_{Z_D}\gg M_Z$, see Sec.~\ref{sec:kmST}.}
    \label{fig:fnlST_AllOs}
\end{figure}

\begin{figure}
    \centering
    \includegraphics[scale=1.0]{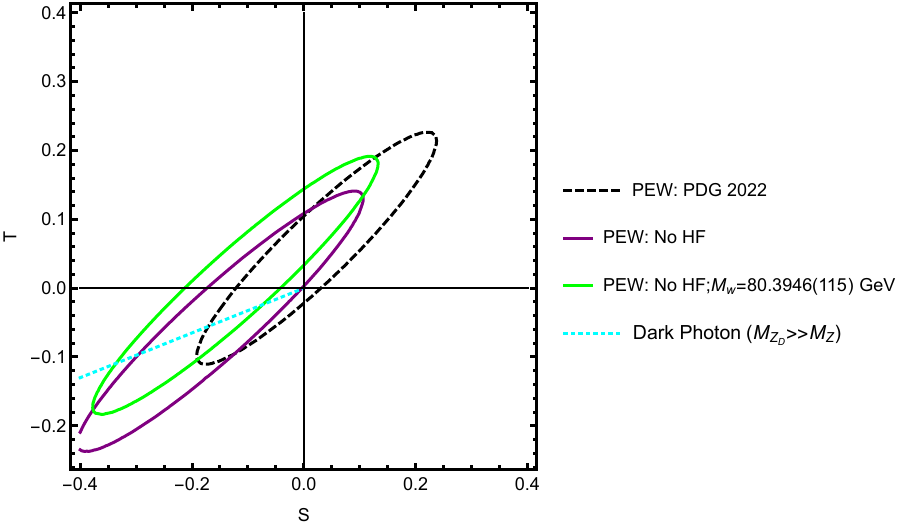}
    \caption{The 2$\sigma$ region in the $S$-$T$ plane (with $U$=0) for different data combinations. The black dashed line is the same as that in Fig.~\ref{fig:fnlST_AllOs} and includes all observables fixed to their 2022 PDG values \cite{ParticleDataGroup:2022pth}.  The purple contour shows the result of omitting the heavy flavor (HF) observables.  The green contour shows the result of omitting the heavy flavor observables and including the CDF measurement in the W Mass average \cite{Amoroso:2023pey}. The dotted cyan line is the line traced out by kinetically mixed dark photon model in the limit $M_{Z_D}\gg M_Z$, see Sec.~\ref{sec:kmST}}.
    \label{fig:fnlST_NoHF}
\end{figure}

Figs.~\ref{fig:fnlST_AllOs} and \ref{fig:fnlST_NoHF} show the preferred regions in the $S$-$T$ plane for different data combinations.  As discussed in Section \ref{sec:oblique}, $U$ is much smaller than $S$ and $T$ in many models of new physics.  We set $U=0$ for our oblique plots, but we allow $U\neq0$ when we explore the dark photon model.  Including $U$ when studying the dark photon ensures the validity of our analysis even when there is not a big hierarchy between the dark photon and SM $Z$ boson masses (see Eqs.~(\ref{eq:STUinKM})).  These plots are produced following the procedures laid out in Section \ref{sec:PEWAnalysis}.  The contours in Fig.~\ref{fig:fnlST_AllOs} use the full collection of measurements laid out in Table \ref{tab:measurements} (omitting $\alpha_s$ for consistency with Refs.~\cite{Haller:2018nnx, Curtin:2014cca}).  The black-dashed contour shows the result of using 2022 PDG \cite{ParticleDataGroup:2022pth} values and is in good agreement with the result from the GFitter group \cite{Haller:2018nnx}.  The point $S = T = 0$ is contained within the contour indicating that the combined fit to the 2022 PDG values is in good agreement with the SM.

To explore the potential impact of a larger $W$-boson mass, we use the combination reported in Ref. \cite{Amoroso:2023pey}, $M_W = 80.3946(115)\ \GeV$.  This combination includes the measurement by the CDF collaboration \cite{CDF:2022hxs}.  The CDF measurement has a low probability of compatibility with the other measurements.  As a result, to include it in the global average, one could also consider schemes where the error bars are inflated.  We explore the impact of this sort of scheme on our analysis by considering a naive average of the CDF, ATLAS, D0, LEP, and LHCb $W$ mass measurements \cite{CDF:2022hxs, ATLAS:2023fsi, D0:2013jba, ALEPH:2013dgf, ParticleDataGroup:2022pth, LHCb:2021bjt} where correlations between the measurements are ignored and the error bars are inflated so that the chi-square per degree of freedom in the combination is equal to one.  Using this prescription we arrive at $M_W = 80.406(16)\ \GeV$.  Exchanging this value for the one above from Ref. \cite{Amoroso:2023pey} would not change our conclusions.
 
The effect of raising $M_W$ is to move the best fit ellipse to larger values of $T$.  See, e.g., Ref.~\cite{Asadi:2022xiy} for a recent analysis of this effect.  This contour is shown in blue in Fig.~\ref{fig:fnlST_AllOs}.  Note that when the CDF measurement is included, the SM point $S = T = 0$ is no longer preferred. 

Next, we discuss the effect of raising $\Delta\alpha^{(5)}_{\text{had}} (M_Z^2)$--as one would do to decrease the tension with the muon $g-2$ measurement.   The value of the hadronic contribution to the running of the fine structure constant, $\Delta\alpha^{(5)}_{\text{had}} (M_Z^2)$, and the hadronic vacuum polarization (HVP) contribution to muon $g-2$ are determined via different dispersion relations from the same cross section $\sigma(e^+e^- \rightarrow \text{hadrons})$ \cite{Keshavarzi:2018mgv, Crivellin:2020zul}.  Therefore, by modifying this cross section one simultaneously shifts both $\Delta\alpha^{(5)}_{\text{had}} (M_Z^2)$ and the HVP.  In Fig.~\ref{fig:fnlST_AllOs} we follow Ref.~\cite{Crivellin:2020zul} and adopt $\Delta\alpha^{(5)}_{\text{had}} (M_Z^2) = .02803(13)$.  This corresponds to making the SM prediction for and the measurement of the anomalous magnetic moment of the muon compatible by changing the cross section at energies below $11.2$ GeV (above this everything is perturbative and better understood) \cite{Crivellin:2020zul}.  As shown in the red contour, the impact of modifying $\Delta\alpha^{(5)}_{\text{had}} (M_Z^2)$ in this way is to push the contour to lower values of $S$.  In agreement with Ref.~\cite{Crivellin:2020zul}, our results show that using modifications in HVP to alleviate the muon $g-2$ tensions worsens the quality of the PEW fit \footnote{As shown in Table \ref{tab:measurements}, our work uses the value of $\Delta\alpha^{(5)}_{\text{had}} (M_Z^2)$ reported in Ref. \cite{ParticleDataGroup:2022pth}. For some recent work on the evaluation of  $\Delta\alpha^{(5)}_{\text{had}} (M_Z^2)$ see Refs. \cite{Narison:2023srj, Erler:2023hyi}.}.  

In Fig.~\ref{fig:fnlST_NoHF}, we again show the ellipse including the full collection of observables at their 2022 PDG values \cite{ParticleDataGroup:2022pth} for comparison purposes.  The purple contour shows the result of  excluding the heavy flavor (HF) observables (the last 6 observables in Table \ref{tab:measurements}) while keeping all other measurements fixed at their 2022 PDG values \cite{ParticleDataGroup:2022pth}.   Due to their correlation, we choose to exclude all the HF observables.  However, the effect is driven by $A_{FB}^b$ - the measurement in most tension with the SM fit.   Excluding the HF observables moves the ellipse towards lower values of both $S$ and $T$. The SM point $S = T = 0$ is no longer preferred (the ellipse only appears to intersect the point $S=T=0$ due to the finite line widths of the contour and the axes).  Similarly, the green contour in Fig.~\ref{fig:fnlST_NoHF} shows the result of excluding HF while simultaneously including the CDF measurement in the average for the W mass following Ref. \cite{Amoroso:2023pey}.  As in Fig.~\ref{fig:fnlST_AllOs}, the effect of raising $M_W$ is to raise $T$.  Again, the SM point $S = T = 0$ is no longer within the preferred region. 

In both Figs.~\ref{fig:fnlST_AllOs} and \ref{fig:fnlST_NoHF} the dotted cyan line shows the line corresponding to the kinetically mixed dark photon in the limit $M_{Z_D}\gg M_Z$ (see Eq.~(\ref{eq:STRatio})).  Moving away from the origin corresponds to increasing the magnitude of the kinetic mixing parameter $\eta$.  The shift induced by excluding the HF observables has the effect of increasing the viable parameter space for the kinetically mixed dark photon.  In fact if  the HF observables are omitted and a higher value of measured $W$ mass is used, a model with a dark photon will be significantly preferred to the Standard Model.

\begin{figure}
    \centering
    \includegraphics[scale=.75]{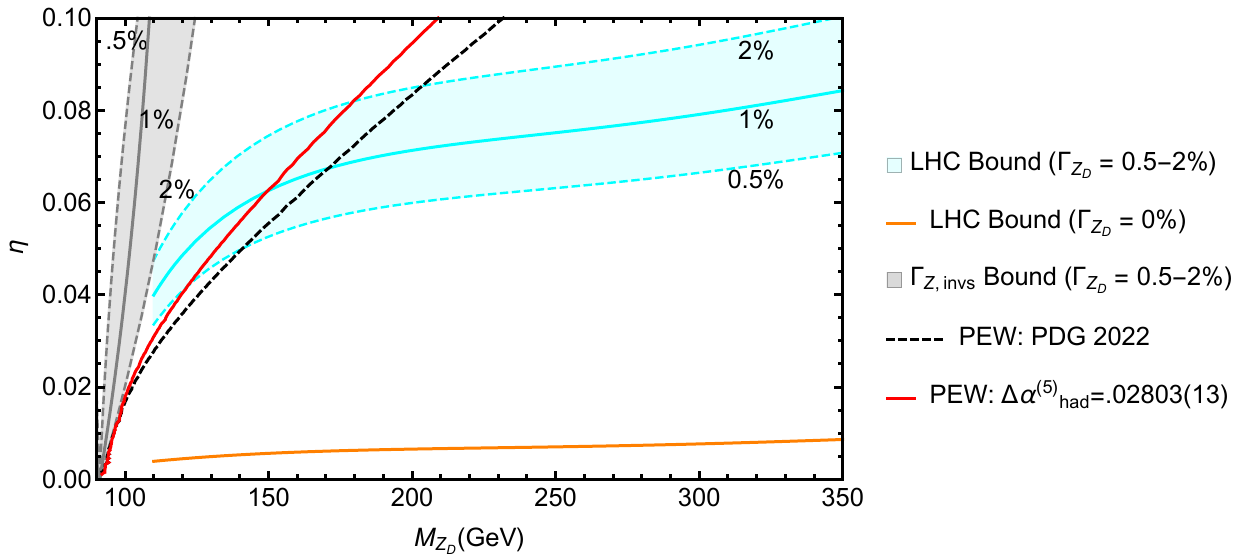}
    \caption{Allowed regions in the $M_{Z_D}-\eta$ plane. The 2$\sigma$ PEW contours are created using Eq.~(\ref{eq:fullPEWDPfit}). The black dashed line shows the upper bound on $\eta$ when using the measurements in the 2022 PDG \cite{ParticleDataGroup:2022pth} in the fit.  The red contour shows how the bound shifts when the hadronic contribution to the running of the fine structure constant is modified.  There is no contour corresponding to the case of only increasing $M_W$ because kinetic mixing in this region of parameter space cannot explain this effect.  The gray shaded region shows how the $Z$ boson invisible width bounds \cite{ALEPH:2005ab, ParticleDataGroup:2022pth} change as the dark photon width $\Gamma_{Z_{D}}$ is varied between 0.5\% and 2\% of $M_{Z_{D}}$. The solid gray line shows our benchmark 1\% fractional width scenario. The orange line is the LHC upper bound on the kinetic mixing parameter if the dark photon has no decays to dark sector states.  The light cyan region shows how these LHC bounds change if the dark photon fractional width is varied between 0.5\% and 2\% with the solid cyan line showing the bound for a 1\%  fractional width.}
    \label{fig:fnlMzPrEta_AllOs}
\end{figure}

\begin{figure}
    \centering
    \includegraphics[scale=0.75]{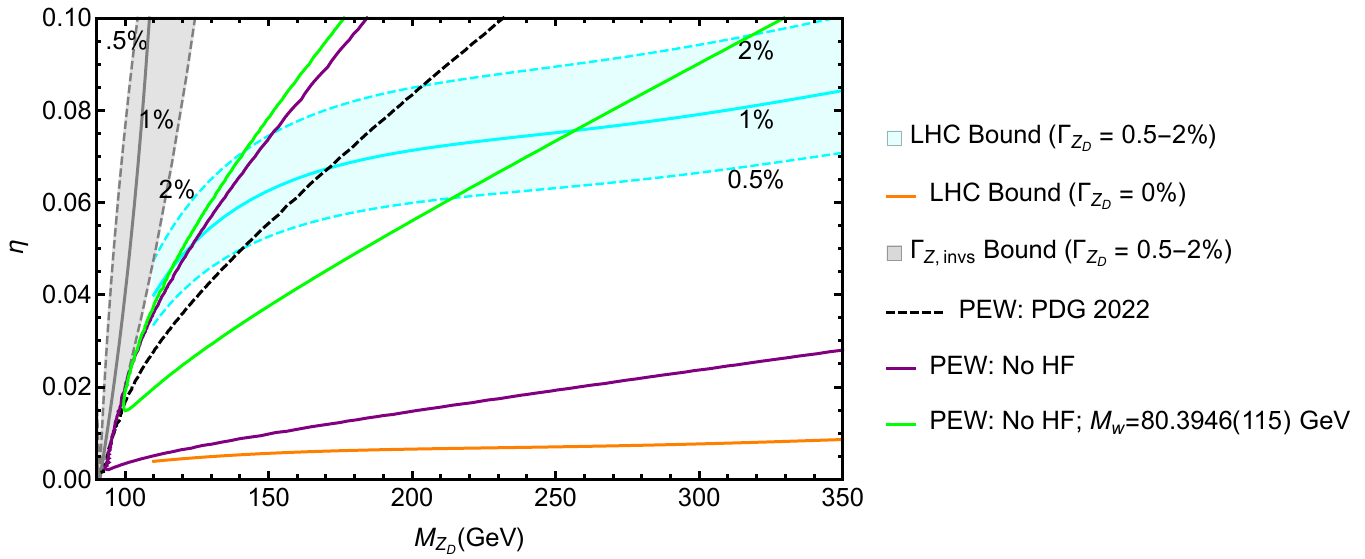}
    \caption{Allowed regions in the $M_{Z_D}-\eta$ plane. The 2$\sigma$ PEW contours are created using Eq.~(\ref{eq:fullPEWDPfit}). The black dashed line shows the PEW upper bound on $\eta$ when the observables take on their 2022 PDG values \cite{ParticleDataGroup:2022pth}. The purple contour shows the result of omitting the heavy flavor (HF) observables.  The green contour shows the result of omitting the heavy flavor observables and including the CDF measurement in the $W$ mass average \cite{Amoroso:2023pey}.  In each of these no-HF cases, the model is constrained to lie between the green/purple lines, i.e., a non-zero $\eta$ is favored. The gray shaded region shows how the $Z$ boson invisible width bounds \cite{ALEPH:2005ab, ParticleDataGroup:2022pth} change as the dark photon fractional width $(\Gamma_{Z_D}/M_{Z_D})$ is varied between 0.5\% and 2\%.  The orange line is the LHC upper bound on the kinetic mixing parameter if the dark photon has no decays to dark sector states.  The light cyan region shows how these LHC bounds change if the dark photon fractional width is varied between 0.5\% and 2\% with the solid cyan line showing the bound for a 1\% fractional width.} 
    \label{fig:fnlMzPrEta_NoHF}
\end{figure}

In Figs.~\ref{fig:fnlMzPrEta_AllOs} and \ref{fig:fnlMzPrEta_NoHF} we use the same data combinations employed in Figs.~\ref{fig:fnlST_AllOs} and \ref{fig:fnlST_NoHF} to place limits on the dark photon model. As mentioned above, to allow for the dark photon to have a mass comparable to that of the SM $Z$ boson, we allow $U$ to be nonzero for this part of the analysis. The translation from the $S-T-U$ parameters space to the $M_{Z_D}-\eta$ parameter space is described in Section \ref{sec:kmST}.
As discussed in that section,  
the region in the $M_{Z_{D}}-\eta$ space is found by requiring 
\begin{equation} \label{eq:fullPEWDPfit}
  \Delta\hat{\chi}^2(M_{Z_D}, \eta) = \hat{\chi}^2(S(M_{Z_D}, \eta), T(M_{Z_D}, \eta), U(M_{Z_D}, \eta)) - \chi^2_{\text{min}} = 6.18.
\end{equation}
This corresponds to the 2$\sigma$ bound for two free parameters, $M_{Z_D}$ and $\eta$.  If instead one imagines fixing a mass for the dark photon, then one would instead only have 1 free parameter and would instead require $\Delta\hat{\chi}^2(M_{Z_D}, \eta) = 3.8$.  This was the route taken by Ref.~\cite{Curtin:2014cca}.  We have verified our procedure against the curve in Ref.~\cite{Curtin:2014cca} and find consistency with their result.  

The black dashed line in Fig.~\ref{fig:fnlMzPrEta_AllOs} is the 2$\sigma$ PEW upper bound on the kinetic mixing parameter when the 2022 PDG values \cite{ParticleDataGroup:2022pth} are used in the fit.  The preferred region of parameter space lies below the black dashed line.  To make this contour, we use the same set of observables that were used to produce the black dashed contours in the $S$-$T$ plots of Figs.~\ref{fig:fnlST_AllOs} and \ref{fig:fnlST_NoHF}.  There is no strong preference for non-zero kinetic mixing, so there is no lower dashed contour for $\eta$. The red contour in Fig.~\ref{fig:fnlMzPrEta_AllOs} corresponds to shifting $\Delta\alpha^{(5)}_{\text{had}} (M_Z^2)$ and uses the same set of observables as the red contour in Fig.~\ref{fig:fnlST_AllOs}.  This weakens the PEW bounds on $\eta$. However, for this value of $\Delta\alpha^{(5)}_{\text{had}} (M_Z^2)$, there is no strong preference for non-zero kinetic mixing. 

For the other data combination shown in Figure~\ref{fig:fnlST_AllOs}, that is, including the CDF $W$ mass measurement while leaving all the other observables unchanged, kinetic mixing in this region of parameter space is unable to completely explain the data.  As a result, there is no contour that corresponds to this scenario in Fig.~\ref{fig:fnlMzPrEta_AllOs}. 

In Fig.~\ref{fig:fnlMzPrEta_AllOs} and \ref{fig:fnlMzPrEta_NoHF}, the orange line shows the existing LHC collider bounds under the assumption of zero width to dark sector states.  The cyan region and dotted and dashed lines shows how the LHC bounds change when a fractional width $(\Gamma_{Z_D}/M_{Z_D})$ of between 0.5\% and 2\% arising from decays to dark sector states is present.  The area above these lines is excluded.  The procedure for adjusting the LHC bound based on the width is described in Section \ref{sec:colliderbounds}.  

The gray lines/region shows the bounds due to the invisible width constraints on the SM $Z$ boson (See Eq.~(\ref{eq:Zinvswid})) for different fractional widths between 0.5\% and 2\%.  The area above the gray lines is excluded. This curve is produced under the assumption that the mass of the final states can be neglected compared to the $Z$ boson and $Z_D$ masses.  Should (all) the dark states to which the $Z_{D}$ decay have masses that are close to (exceed) $M_{Z}/2$, these bounds can by modified (absent).  Because these bounds are most relevant for $M_{Z_{D}}$ close to $M_{Z}$, it would require a coincidence for the dark states to be accessible in the decay of one gauge boson but not the other.  Between $75$ and $110$ GeV, there are no LHC bounds on $Z_{D}$ production.

For dark photons with width to a dark sector, the PEW and collider bounds are complementary.  Consider the case of a $Z_{D}$ with a 1\% fractional width. Between 110 and $\sim170$ GeV the PEW constraint (using the 2022 PDG measurements \cite{ParticleDataGroup:2022pth}) is stronger than the collider constraint by a factor of $\sim1.25$ and, as alluded to above,  below 110 GeV there is no LHC constraint. In the region below 110 GeV it might be possible that a $Z_{D}$ that decays exclusively to the SM could be hiding in the LHC data, though it seems unlikely that a narrow resonance would have escaped notice. A very wide ($\Gamma_{Z_{D}}/M_{Z_{D}} \sim {\mathcal O} (10 \%)$) $Z$-prime in this region would be especially challenging to search for owing to the large SM $Z$ background.  However, the bound from the invisible width of the SM $Z$ boson (see grey line of Figs.~\ref{fig:fnlMzPrEta_AllOs},\ref{fig:fnlMzPrEta_NoHF})  becomes increasingly strong as the $Z_{D}$ gets wider, and  naively disfavors such a wide $Z_{D}$ in the $<110$ GeV range -- but as a caveat, if the mass of the $Z_{D}$ is sufficiently close to the $Z$ boson mass, this could bias the extraction  of the $Z$ parameters, including bounds on the invisible width, see Appendix \ref{sec:nonoblique}.  In any case, such a wide $Z_{D}$ could be present at higher masses.

In Fig.~\ref{fig:fnlMzPrEta_NoHF}, we show the impact of excluding the heavy flavour observables in the PEW fit, for both the case of the 2022 PDG $W$ mass (purple), and the case where the larger CDF $W$ mass is averaged in (green).  The preferred region is between the purple (green) lines.  In both No-HF scenarios, non-zero KM is preferred (i.e., the line $\eta = 0$ is not contained in the preferred regions).  When the CDF measurement is averaged into the value of the $W$ mass and the HF observables are excluded, the preference for non-zero kinetic mixing is strong.  Interestingly, in this scenario, between 110 and $\sim250$ GeV there is viable parameter space below the blue LHC line(s) and within the green region.  That is, the corresponding set of PEW measurements would favor a $Z_{D}$ with parameters not yet directly probed at the LHC.  To  exclude this scenario, the collider bounds on $\eta$ would only need to improve by a factor of $\sim 2$.  This represents an interesting near-term target for LHC searches. Of course, for this parameter space to be truly allowed would require an understanding of the discrepancy of the HF data with the other PEW measurements. If due to other new physics, then this should be included in the fit  as well, which could change the allowed parameter space.  If due to an unknown systematic error, then the noted parameter space is of particular interest.

Furthermore, in the no HF scenario with PDG $W$ mass, between $\sim 95$ and $110$ GeV non-zero KM is preferred, the invisible width bounds are relatively weak, and there is no LHC bound.  Our results support the case for a dedicated LHC search for a kinetically mixed dark photon in the region on and around the $Z$ pole.  The data for this region already exists and performing a dedicated search would allow the discovery or exclusion of interesting possibilities allowed by the PEW fit. 

\section{Conclusion}\label{sec:conclusion}

In this work, we studied how tensions involving the mass of the $W$ boson, the hadronic contribution to the running of the fine structure constant, and the heavy-flavor observables can affect the precision electroweak fit.  Figs.~\ref{fig:fnlST_AllOs} and \ref{fig:fnlST_NoHF} show how the preferred region in the $S-T$ plane changes when these tensions are dealt with in different ways.  Summarizing,  raising $M_W$ favors higher values of $T$, raising $\Delta\alpha^{(5)}_{\text{had}}$ favors slightly lower $S$ values, and excluding the heavy flavor observables favors smaller $S$ and $T$ would make the data inconsistent with the SM point $S = T = 0$.

Next, we expanded our oblique analysis to allow for nonzero $U$ and applied the results to the dark photon model (see Appendix \ref{sec:nonoblique} for discussion of when oblique analysis is insufficient).  We compute the preferred dark photon parameter space for different data combinations.  The results are presented in Figs.~\ref{fig:fnlMzPrEta_AllOs} and \ref{fig:fnlMzPrEta_NoHF}.  In the no heavy flavor scenarios, non-zero kinetic mixing is preferred.  

If the $Z_{D}$ decays exclusively to the Standard Model, direct searches at the LHC are generally a much stronger probe than precision electroweak constraints.  However, in the well-motivated case of that there are light dark sector states to which the $Z_{D}$ has a significant branching ratio, we have shown that either direct searches or PEW constraints can be the most sensitive probe.  Moreover, if any of the tensions in the PEW fit were due to the presence of $Z_{D}$, the preferred value of the kinetic mixing would be close to the naive PEW bounds.   There is strong motivation to continue improving the search bounds for $Z_{D} \rightarrow \ell \ell$ in this case.  Our results also support a dedicated LHC search in the region between $M_Z$ and $110$ GeV that was excluded from previous analyses.

\acknowledgments

EP would like to acknowledge useful conversations with Prudhvi Bhattiprolu, Cristina Mantilla Suarez, James Wells, Keith Riles, Yik Chuen San, and Jianming Qian.
AP would like to thank Chris Hayes. 
KH~was partly supported by Grant-in-Aid for Scientific Research from the Ministry of Education, Culture, Sports, Science, and Technology (MEXT), Japan 20H01895 and by World Premier International Research Center Initiative (WPI), MEXT, Japan (Kavli IPMU).
EP was supported in part by a Leinweber Graduate Summer Fellowship. AP was supported by DoE grant DE-SC0007859. This work made use of \textsc{webplotdigitizer}\footnote{\url{https://automeris.io/WebPlotDigitizer/}}, \textsc{numpy} \cite{Harris:2020xlr}, \textsc{scipy} \cite{Virtanen:2019joe}, and \textsc{matplotlib} \cite{Hunter:2007ouj}.

\appendix
\section{Non-Oblique Corrections}
\label{sec:nonoblique}

The oblique corrections capture effects due to the modification of the $Z$ boson mass and its couplings to SM fermions.  They do not capture the effects of $Z-Z_D$ interference nor the resonant production of $Z_D$ and subsequent decays to SM fermions.  These effects are subdominant for data taken at the $Z$ pole, unless the mass of the dark photon is nearly degenerate with that of the SM $Z$ boson.  In our discussion of precision electroweak constraints, we have tacitly assumed that these effects are indeed small.   Here we explore how and when these effects can become significant.

We highlight two arenas where these non-oblique corrections could play a role.  The first is at the level of the traditional PEW fit.  If the non-oblique contributions are significant, then they will modify the theoretical expressions for the $Z$-peak pseudo-observables ($M_Z, \Gamma_Z, \sigma^0_{\text{had}}$, etc.).  To fully take into account this modification, one would need to compute the full theoretical prediction for these observables in the context of the dark photon model and redo the fit.

\begin{figure}
    \centering
    \includegraphics[scale=0.85]{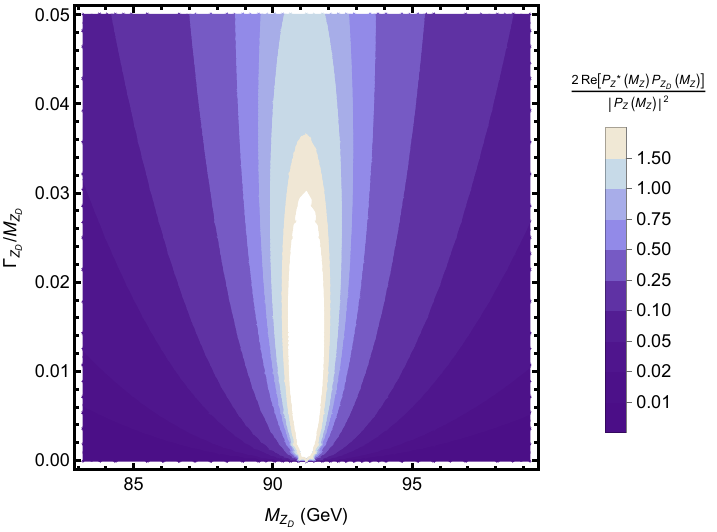}
    \caption{Ratio of the $Z-Z_D$ interference propagator combination to the $Z$ boson propagator squared for dark photon masses near the mass of the SM $Z$ boson.  The oblique corrections go like $\sim \eta^2 |P_Z(M_Z)|^2$ while the non-oblique corrections go like $\sim \eta^2 2 \text{Re}[P_Z(M_Z)^{\ast}P_{Z_D}(M_Z)]$.  We see that the non-oblique corrections are about 2 orders of magnitude below the oblique corrections for masses above $\sim95$ GeV and fractional widths below $\sim 2\%$.}
    \label{fig:IntZSqRatio}
\end{figure}

To estimate where this effect becomes significant, we compare the magnitude of the contributions around the $Z$ pole.  Let $P_X(q) = 1/(q^2 - M_X^2 + iM_X\Gamma_X)$ denote the propagator for a particle $X$ with mass $M_X$ and width $\Gamma_X$ evaluated at energy scale $q$.  The oblique parameters capture corrections that go like $\sim \eta^2 |P_Z(M_Z)|^2$ and the leading order non-oblique effects go like $\sim \eta^2 2 \text{Re}[P_Z(M_Z)^{\ast}P_{Z_D}(M_Z)]$. The resonant production of $Z_D$ is suppressed by another factor of $\eta^2$ and is never dominant at the $Z$ pole for reasonable values of $\eta$.  Fig.~\ref{fig:IntZSqRatio} shows the result of comparing these two contributions.  We see that for the parameter space considered in this work (fractional widths below 2\% and masses $\gtrsim 95$ GeV), the oblique corrections are expected to be be $2$ or more orders of magnitude larger than the non-oblique corrections for data taken at the $Z$ pole. 

The second place where the non-oblique corrections could be significant is at the level of the initial inference of the $Z$-peak pseudo-observables ($M_Z, \Gamma_Z, \sigma^0_{\text{had}}$, etc.).  These pseudo-observables are themselves determined from fits to data, and it is possible that the presence of $Z-Z_D$ interference could effect the fit.  For example, the $Z$-boson mass, width, and total hadronic cross section pseudo-observables are found by fitting a Breit-Wigner curve to the observed $Z$-boson peak.  We emphasize that while the usual approach allows for modified couplings of the $Z$, it does not allow for a new interference term or a second resonance in close proximity to the $Z$ boson \cite{ALEPH:2005ab}.  If there is a dark photon with a mass very close to that of the SM $Z$ boson, this could bias this fit, i.e., lead one to infer different values of these pseudo-observables. 

While it is difficult to quantitatively explore this effect without a thorough understanding of systematics, here we make some initial explorations. We start with the $Z$-pole data for $e^+e^- \rightarrow \text{hadrons}$ at the DELPHI experiment \cite{DELPHI:2000wje}.  The data consists of 10 data points taken at the $Z$ peak and at $\sim 1.8$ GeV above and below the peak.  We then calculate the tree-level SM cross section for $e^+e^- \rightarrow \text{hadrons}$, $\sigma_{SM}(s)$.  Then, fixing the $Z$-boson mass and couplings to SM fermions, we compute a cross section that incorporates the contribution of kinetic mixing (KM): $e^+e^- \rightarrow \text{hadrons}$, $\sigma_{KM}(s, \eta, M_{Z_D}, \Gamma_{Z_D})$.  This KM cross section is the SM one plus the effects of $Z-Z_D$ interference and resonant $Z_D$ production.   Explicitly, we compute the squared amplitude, 
\begin{equation}
    |\mathcal{M}(e^+e^- \rightarrow \text{hadrons})|^2 = |\mathcal{M}_{\gamma} + \mathcal{M}_{Z, SM} + \mathcal{M}_{Z_D}|^2
\end{equation}
where $\mathcal{M}_{\gamma}$ and $\mathcal{M}_{Z, SM}$ are the amplitudes computed exactly as they would be in the SM.

Then, for a chosen set of $\eta,  M_{Z_D}, \Gamma_{Z_D}$, we generate a simulated data sets for both the KM and SM scenarios.  To do this, we evaluate the cross sections at the $\sqrt{s}$ values provided by DELPHI.\footnote{Due to radiation effects, the peak of the DELPHI data is not exactly on the $Z$ pole.  We shift the $\sqrt{s}$ values so that they are centered on the $Z$ pole.}  For each $\sqrt{s}$ we generate a mock data point by drawing from a Gaussian distribution with mean $\mu = \sigma_{SM}$ ($\mu = \sigma_{KM}$ for the KM case) and standard deviation given by the reported DELPHI error at that value of $\sqrt{s}$.  We generate 1000 mock data sets for each choice of $\eta,  M_{Z_D}, \Gamma_{Z_D}$.  For each data set, we fit a Breit-Wigner curve (including photon and $\gamma$-$Z$ interference contributions) and infer values of $M_Z$, $\Gamma_Z$, and $\sigma_{\text{had}}^0$.  Then, we average over the 1000 mock data sets to get two sets of inferred values of these parameters - one in the case of KM and one in the case of the SM.

\begin{figure}
    \centering
    \includegraphics[scale=0.85]{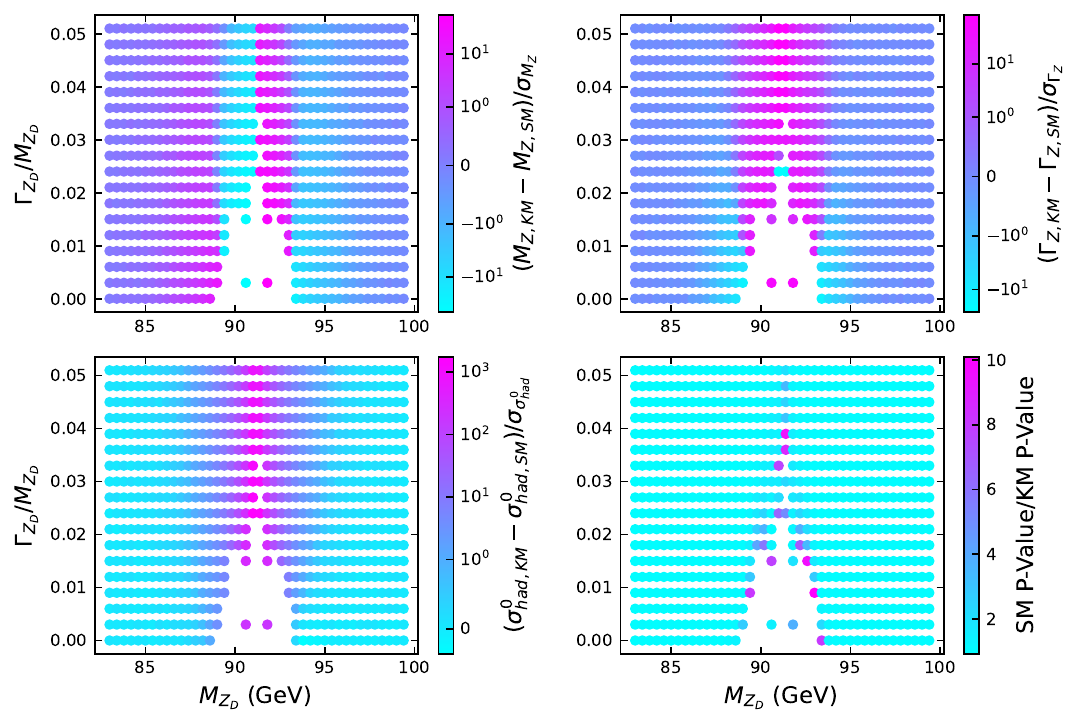}
    \caption{Shift in the inferred values of $M_Z$, $\Gamma_Z$, and $\sigma_{\text{had}}^0$ and change in the p-value of the $Z$ resonance peak fit in the presence of a kinetically mixed dark photon with $\eta=0.01$.  The change is shown in units of the parameter uncertainty as inferred from the fit.  Points where the p-value of the fits exceed 0.05 are not shown.}
    \label{fig:ZPeakShifts}
\end{figure}

Fig.~\ref{fig:ZPeakShifts} shows the result of this simulation for $\eta = 0.01$.  The $Z$ boson mass and $Z$ couplings to SM fermions are identical in the two scenarios.  However, if kinetic mixing is present, the inferred values of $M_Z$, $\Gamma_Z$, and $\sigma_{\text{had}}^0$ could be shifted either higher or lower depending on the mass and width of the dark photon. In some cases, particularly when $M_{Z_D}$ is close to $M_Z$ and the width is very narrow, the Breit-Wigner curve (plus $\gamma$ and $\gamma-Z$ interference) provides a very poor fit.  In these cases, it would be easy to tell that there is another particle.  In Fig.~\ref{fig:ZPeakShifts}, we have only shown points where the p-value of the both fits is greater than 0.05. 

The new physics contributions for different values of $\eta$ scales like $\eta^2$.  The magnitude of the shifts in the inferred parameters also changes with $\eta$, but the scaling is not straightforward.  We repeat the analysis for values of $\eta$ up to $0.06$ and examine where the shifts in any of the inferred parameters exceed $1\sigma$.  The region where the shifts become significant (i.e., at least one inferred pseudo-observable is shifted by $> 1\sigma$) is in an area of $M_{Z_D}-\eta$ parameter space that is already excluded by the invisible width bounds. 

One interesting subtlety is that shifts in the inferred pseudo-observables could also shift the invisible width bound.  The invisible width of the $Z$ boson can be determined from \cite{ALEPH:2005ab}.
\begin{equation}
    R^0_{\text{inv}} = \left(\frac{12\pi R_{\ell}^0}{\sigma^0_{\text{had}}M_Z^2}\right)^{1/2} - R_{\ell}^0 - (3 + \delta_{\tau})
\end{equation}
where $R^0_{\text{inv}} \equiv \Gamma_{\text{inv}}/\Gamma_{\ell\ell}$, $R_{\ell}^0 \equiv \Gamma_{\text{had}}/\Gamma_{\ell\ell}$, $\Gamma_{\ell\ell}$ is the partial width of $Z$ to massless leptons, and $\delta_{\tau}$ accounts for the difference between $\Gamma_{\ell\ell}$ and $\Gamma_{\tau\tau}$ due to the mass of the tau lepton.
If $M_Z$ or $\sigma_{\text{had}}^0$ is overestimated (so the actual value is lower), $R^0_{\text{inv}}$ inferred from the above equation becomes smaller than the actual value, potentially relaxing the invisible width bound.
Above the $Z$ pole in Fig.~\ref{fig:ZPeakShifts}, $M_Z$ and $\sigma_{\text{had}}^0$ are shifted in opposite directions, and this effect would partly cancel.

The analysis of this section is designed to get an idea of where $Z-Z_D$ interference effects become relevant.  Our analysis suggests the following takeaways: 1) For $M_{Z_D} \gtrsim 110$ GeV, it is safe to ignore $Z-Z_D$ interference effects for the range of $\eta$ values considered in this work, and for $100 \lesssim M_{Z_D} \lesssim 110$ GeV the interference effects can be neglected provided that $\eta \lesssim 0.03$. 2) Below $100$ GeV, more robust analysis is required to understand the interplay between the invisible width bounds and the pseudo-observable biasing due to $Z-Z_D$ interference.  We leave this investigation to future work. 

\bibliographystyle{jhep}

\bibliography{NewBibliography}

\end{document}